\documentclass[twocolumn]{emulateapj} 
\usepackage{natbib}
\usepackage{epsfig}
\usepackage{graphicx} 
\usepackage{subfigure}
\usepackage{float}
\usepackage{amsmath}
\usepackage{color}
\usepackage{amssymb}
\usepackage{amsfonts}
\usepackage{units} 
\usepackage{mathtools}
\usepackage{bm}
\usepackage[colorlinks,linkcolor=blue,anchorcolor=green,citecolor=blue]{hyperref}
\def\be{\begin{eqnarray}}
\def\ee{\end{eqnarray}} 

\shorttitle{Test of special relativity via Doppler effect}
\shortauthors{Yang, Zhu \& Zhang} 
\begin{document}

\title{Relativistic Astronomy. III. test of special relativity via Doppler effect}

\author{Yuan-Pei Yang\altaffilmark{1}, Jin-Ping Zhu\altaffilmark{2,3} and Bing Zhang\altaffilmark{4}}

\affil{
$^1$ South-Western Institute for Astronomy Research, Yunnan University, Kunming, Yunnan, P.R.China; ypyang@ynu.edu.cn;\\
$^2$Kavli Institute for Astronomy and Astrophysics, Peking University, Beijing 100871, China;\\
$^3$ Department of Astronomy, School of Physics, Peking University, Beijing 100871, China \\
$^4$ Department of Physics and Astronomy, University of Nevada, Las Vegas, NV 89154, USA; zhang@physics.unlv.edu
}

\begin{abstract}
The ``Breakthrough Starshot'' program is planning to send transrelativistic probes to travel to nearby stellar systems within decades. Since the probe velocity is designed to be a good fraction of the light speed, \citet{zha18} recently proposed that these transrelativistic probes can be used to study astronomical objects and to test special relativity. In this work, we further propose some methods to test special relativity and constrain photon mass using the Doppler effect with the images and spectral features of astronomical objects as observed in the transrelativistic probes. We introduce more general theories to set up the framework of testing special relativity, including a parametric general Doppler effect and Doppler effect with massive photon. We find that by comparing the spectra of a certain astronomical object, one can test Lorentz invariance and constrain photon mass. Besides, using imaging and spectrograph capabilities of transrelativistic probes, one can test time dilation and constrain photon mass. For a transrelativistic probe with velocity $v\sim0.2c$, aperture $D\sim3.5~\unit{cm}$ and spectral resolution $R\sim100$ (or $1000$), we find that the probe velocity uncertainty can be constrained to $\sigma_v\sim0.01c$ (or $0.001c$), and the time dilation factor uncertainty can be constrained to $\Delta\gamma=|\hat\gamma-\gamma|\lesssim0.01$ (or $0.001$), where $\hat\gamma$ is the time dilation factor and $\gamma$ is the Lorentz factor. Meanwhile, the photon mass limit is set to $m_\gamma\lesssim10^{-33}~\unit{g}$, which is slightly lower than the energy of the optical photon.
\end{abstract}


\section{Introduction}

The ``Breakthrough Starshot'' program aims at proving the concept of developing an unmanned space probe traveling at a good fraction of light speed, which is designated to travel to nearby stellar systems within decades. 
``Sprites'' as the first prototype of the Breakthrough Starshot program has recently been launched at a low-Earth orbit\footnote{https://breakthroughinitiatives.org/News/12}. It consists of $3.5~\unit{cm}\times3.5~\unit{cm}$ chips that weigh just 4 grams each.
Recently, \citet[][Paper I]{zha18} suggested that using such a transrelativistic probe to observe celestial objects, one may study the astronomical objects in a unique way and perform tests on fundamental physics. In order to carry out such tasks, the motion of the probe (velocity and direction of motion) needs to be solved. 
Through detailed simulations, \citet[][Paper II]{zhu18} showed that by comparing the positions of at least three point sources in the Earth frame and in the probe frame, the probe motion velocity and direction can be fully solved within the framework of special relativity. Introducing another point source allows test of special relativity light aberration. An upper limit of photon mass can be also derived from the non-deviation of the aberration angle under the de Broglie-Proca theory \citep{pro36,deb40}.

In this work, we discuss another method to test special relativity and constrain photon mass, making use of the Doppler effect (see also \citealt{zha18}, Section 2.4). 
This paper is organized as follows: In Section \ref{zhu}, based on \citet{zhu18}, we discuss the Doppler effect in the transrelativistic probe frame and its corresponding uncertainty. Next, we introduced more generalized theories to set up the framework of testing special relativity via the Doppler effect (Section \ref{theory}). We define a \emph{parametric, generalized Doppler effect} in \ref{DE}. The Doppler effect for massive photon under the de Broglie-Proca theory \citep{pro36,deb40} is discussed in \ref{photonmass}. In \ref{inv}, we discuss the Doppler transformations beyond special relativity. In Section \ref{lisf}, we develop a method to test the Lorentz invariance by comparing the spectra of a certain astronomical object in different frames. In Section \ref{slda}, we develop another method to test time dilation and photon mass via spectral lines and directional angles. The results are summarized and discussed in Section \ref{discon}.

\section{Doppler effect on a transrelativistic probe in special relativity}\label{zhu}

\cite{zhu18} proposed that one can measure the probe motion velocity and direction by comparing the positions of three point sources in the Earth frame and in the probe frame within the framework of special relativity. Here we assume that the probe motion velocity and direction have been obtained using this method. We then select an astronomical object whose observation direction and spectral features can be measured in the transrelativistic probe frame to test the Doppler effect. 

We define that the probe motion velocity is $v$ and the directional angle of the selected object related to the probe motion direction is $\theta'$ in the probe frame. We identify a spectral line from the selected object whose wavelength is $\lambda'$ and $\lambda$ in the probe frame and the Earth frame, respectively\footnote{This line does not have to have the frequency in the lab. It can be a cosmologically redshifted line. It is the difference between the wavelengths (frequencies) in the two frames (Earth frame and probe frame) that matters.}. In the probe rest frame, the theoretical Doppler factor based on special relativity is 
\be
\mathcal{D}_{\rm the}=\frac{1}{\gamma(1-\beta\cos\theta')},\label{dthe}
\ee
where $\gamma=1/\sqrt{1-\beta^2}$ and $\beta=v/c$.
On the other hand, by measuring the spectral line wavelengths in different frames, one can obtain the observed Doppler factor
\be
\mathcal{D}_{\rm obs}=\frac{\lambda}{\lambda'}.\label{dobs}
\ee
Comparing $\mathcal{D}_{\rm the}$ and $\mathcal{D}_{\rm obs}$, the Doppler effect can be tested. For example, we take $v=0.2c$, $\theta'=\pi/6$, and $\lambda'=400~\unit{nm}$ as the initial input values. According to the aberration method discussed in \citet{zhu18}, the uncertainty of the probe motion velocity is $\sigma_\beta\sim10^{-5}$ and the uncertainty of the probe motion direction is $\sigma_{\theta'}\sim\lambda'/D\sim10^{-5}$, where the probe aperture for the Breakthrough Starshot is taken as $D\sim3.5~\unit{cm}$. Most interference-filter imaging has an spectral resolution $R\sim\lambda/\delta\lambda\sim(100-10000)$.
Considering that the probe is designed to be small, here we take a fiducial spectral resolution $R\sim(100-1000)$ and get $\mathcal{D}_{\rm the}\simeq\mathcal{D}_{\rm obs}\simeq1.18505$ with $\sigma_{\mathcal{D}_{\rm the}}\sim10^{-5}$ and $\sigma_{\mathcal{D}_{\rm obs}}\sim10^{-2}$ for $R\sim100$ ($\sigma_{\mathcal{D}_{\rm obs}}\sim10^{-3}$ for $R\sim1000$). Since $\sigma_{\mathcal{D}_{\rm the}}\ll\mathcal{D}_{\rm the}$ and $\sigma_{\mathcal{D}_{\rm obs}}\ll\mathcal{D}_{\rm obs}$, the Doppler effect can be well tested in the transrelativistic probe within the framework of special relativity to the precision of the spectrograph.

Besides testing the Doppler effect under the framework of special relativity, one can also test special relativity itself and constrain the mass of photon within more generalized theories. The corresponding methods are discussed below in Section \ref{theory}, and the tests are discussed in Sections \ref{lisf} and \ref{slda}. 

\section{Doppler effect on a transrelativistic probe in generalized theories}\label{theory}

\subsection{A Parametric Generalized Doppler effect}\label{DE}

Testing special relativity is usually carried out within the Robertson-Mansouri-Sexl (RMS) framework \citep{rob49,man77}.
This framework assumes
\begin{eqnarray}
 t' & = & a(v) t+\epsilon(v) x', \nonumber\\
 x' & = & b(v)(x-vt),\nonumber\\
 y' & = & d(v)y,\nonumber\\
 z' & = & d(v)z.\label{rms}
\end{eqnarray}
where $x,y,z,t$ are measured in a postulated preferred frame, and $x',y',z',t'$ are measured in a comoving frame with velocity $v$ along $x'$ direction related to the preferred frame. $a(v),b(v),d(v),\epsilon(v)$ are functions as relative velocity.
In particular, to the second order in $v/c$, a velocity-independent parameter $\alpha$ of the RMS framework has the following form 
\be
a(v)\simeq1+\alpha v^2/c^2.
\ee
For special relativity, one has $1/a(v)=b(v)=\gamma\equiv1/\sqrt{1-v^2/c^2}$, $d(v)=1$, $\epsilon(v)=-v/c^2$, and $\alpha=-1/2$. 
On the other hand, the time dilation effect is described by the following \emph{assumption}:

\begin{itemize}

\item \emph{Assumption}: Considering two frames $K$ and $K'$ with a relative velocity $v$, for a clock at rest in the comoving frame $K'$, 
the relationship between the time intervals measured in Frames $K$ and $K'$, respectively, is
\be
\Delta t=\hat\gamma(v)\Delta t',\label{gamma}
\ee
where $\hat\gamma(v)$ is defined as the time dilation factor that depends on the relative velocity $v$. 

\end{itemize}
Thus, under the RMS theory, one has $a=1/\hat\gamma$ and $\alpha+1/2=(1/\hat\gamma-1/\gamma)/\beta^2$, where $\beta=v^2/c^2$.
For Galileo transformation, one always has $\hat\gamma(v)=1$. For Lorentz transformation, one has $\hat\gamma(v)=\gamma\equiv1/\sqrt{1-v^2/c^2}$, where $\gamma$ is the Lorentz factor in special relativity.

\begin{figure}
\centering
\includegraphics[angle=0,scale=0.3]{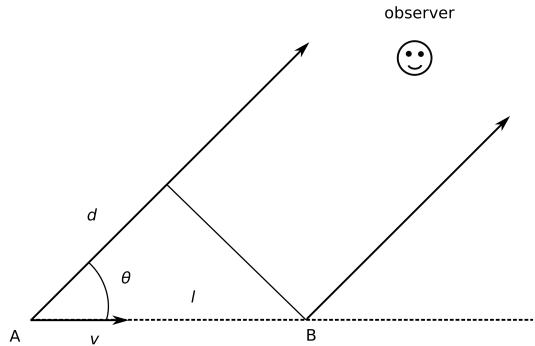}
\caption{Geometry for the generalized Doppler effect.}\label{fig1}
\end{figure}

Based on \emph{Assumption}, we discuss a \emph{generalized} Doppler effect with the parameter $\hat\gamma$. We consider that the source is moving and emitting electromagnetic waves, as shown in Figure \ref{fig1}. In the observer rest frame $K$, the moving source emits radiation with one period as it moves from Point A to Point B with the velocity $v$. Define the radiation frequency in the source rest frame $K'$ as $\nu'$. According to \emph{Assumption}, e.g., Eq.(\ref{gamma}), the time it takes to move from Point A to Point B in Frame $K$ is
\be
\Delta t_e = \frac{\hat\gamma}{\nu'}.
\ee
As shown in Figure \ref{fig1}, one has $l=v\Delta t$ and $d=l\cos\theta$. The difference in the arrival times $\Delta t_a$ of the radiation emitted at Point A and Point B is equal to $\Delta t$ minus the time taken for the radiation to propagate across the distance $d$. One thus has
\be
\Delta t_a = \Delta t_e-\frac{d}{c_\gamma}=\Delta t_e\left(1-\beta\cos\theta\right),
\ee
where $\beta= v/c_\gamma$ is the dimensionless velocity of the moving source. Note that here $c_\gamma$ is emphasized to be the light velocity in the observer frame $K$. For special relativity, one has $c_\gamma=c$, where $c$ is defined as the ultimate maximum velocity.
Therefore, the observed frequency is
\be
\nu=\frac{1}{\Delta t_a}=\frac{\nu'}{\hat\gamma(1-\beta\cos\theta)}\equiv\hat{\mathcal{D}}\nu',\label{doppler}
\ee
where $\hat{\mathcal{D}}\equiv {1}/{[\hat\gamma(1-\beta\cos\theta)]}$.

In general, for a moving probe with a camera, one usually defines the probe frame as $K'$ and the Earth rest frame as $K$. \citet{zha18} and \citet{zhu18} adopted this convention, which we will keep in the following discussion. The Doppler transformation for frequencies can be then written as
\be
\nu'=\frac{\nu}{\hat\gamma(1-\beta\cos\theta')}\equiv\hat{\mathcal{D}}\nu,\label{doppler}
\ee
where 
\be
\hat{\mathcal{D}}\equiv \frac{1}{\hat\gamma(1-\beta\cos\theta')}.
\ee
is the \emph{generalized Doppler factor}, $\beta=v/c_{\gamma}'$, $c_{\gamma}'$ is the light speed, and $\theta'$ is the source directional angle in the probe frame $K'$. Due to Eq.(\ref{doppler}), one has the time transformation relation $dt'=\hat{\mathcal{D}}^{-1}dt$.
Note that Eq.(\ref{doppler}) has the same form as that of special relativity \citep{ryb79}, with the only difference being replacing the Lorentz factor $\gamma$ by the time dilation factor $\hat\gamma$.

\subsection{Doppler effect for massive photon in special relativity}\label{photonmass}

Next, we discuss the Doppler effect for massive photon under the de Broglie-Proca theory. de Broglie-Proca theory is used to describe electrodynamics with non-zero photon mass \citep{pro36,deb40}. Under the de Broglie-Proca theory, the physical constant $c$ corresponds to the ultimate maximum speed rather than light speed, and Lorentz invariance remains valid. Therefore, one can say that de Broglie-Proca theory is still within the framework of special relativity, but light speed is no longer constant because of the photon mass. 

According to the energy-momentum relation, i.e.,
\be
E=h\nu=\sqrt{p^2c^2+m_\gamma^2c^4},
\ee
the photon velocity satisfies \citep[e.g.][]{deb40,yan17,zhu18}
\be
c_\gamma(\nu)=\frac{\partial E}{\partial p}=c\sqrt{1-\left(\frac{m_\gamma c^2}{h\nu}\right)^2}.\label{vm}
\ee
Similar to the discussion in Section \ref{DE}, the frequency Doppler transformation is given by
\be
\nu'=\frac{\nu}{\gamma(1-\beta_m(\nu')\cos\theta')}\equiv\mathcal{D}_m(\nu')\nu,\label{dopm}
\ee
where $\mathcal{D}_m(\nu')\equiv1/[\gamma(1-\beta_m(\nu')\cos\theta')]$ is defined as the Doppler factor for massive photon, $\beta_m(\nu')=v/c_\gamma'\equiv v/c_\gamma(\nu')$ is the dimensionless velocity of photon at frequency $\nu'$, and $\gamma=1/\sqrt{1-\beta^2}$ is the Lorentz factor with $\beta=v/c$, and $c$ is the ultimate maximum speed. Note that for massive photon, the Doppler factor depends on the photon frequency.

Since Lorentz invariance remains valid within the de Broglie-Proca theory, one has invariant four-volume $dtd^3\bm{x}=dtdV$, invariant phase-space element $d^3\bm{p}/E$ and invariant phase volume $d\mathcal{V}=d^3\bm{x}d^3\bm{p}$. Therefore, other Doppler transformations, e.g, volume ($dV'=\mathcal{D}_mdV$), intensity ($j_{\nu'}'=\mathcal{D}_m^2j_{\nu}$) etc., also keep the same forms as in special relativity, with $\mathcal{D}_m$ replaced by $\mathcal{D}$ \citep[e.g.][]{ryb79,der09}.

\subsection{Doppler transformations in more generalized theory}\label{inv}

As discussed in Section \ref{DE}, the generalized Doppler transformation for frequency, i.e., $\nu'=\hat{\mathcal{D}}\nu$, (and the corresponding transformation in time, i.e., $dt'=\hat{\mathcal{D}}^{-1}dt$) has the same form as that in special relativity, with $\hat\gamma$ taking a more general form. However, for some more generalized Doppler transformations, such as volume transformation ($dV'=\hat{\mathcal{D}}dV$), intensity transformation ($j_{\nu'}'=\hat{\mathcal{D}}^2j_{\nu}$), the transformation forms may not only depend on the Doppler fact $\hat{\mathcal{D}}$. The reason is that the Lorentz invariance may break beyond special relativity. 

For example, let us consider volume transformation. In general, the transformation of four-volume $dtdV=dtd^3\bm{x}$ between two frames is given by $dt'dV'=dt'd^3\bm{x}'=J(v)dtd^3\bm{x}=J(v)dtdV$, where $J(v)$ is the Jacobian of the transformation. The Jacobian always satisfies $J(v)=1$ for special relativity \citep{der09}, leading to $dt'dV'=dtdV$ and $dV'=\hat{\mathcal{D}}dV$. 
However, in the RMS framework, the Jacobian $J(v)$ depends on the explicit form of
the frame transformation, e.g., Eq.(\ref{rms}), and $J(v)=1$ is allowed to be not satisfied. Thus, in this case the Jacobian $J(v)$ depends on the RMS parameters $a(v),b(v),d(v),\epsilon(v)$, and is a function of the relative velocity $v$. 
According to Eq.(\ref{doppler}), the volume transformation is 
\be
dV'=J(v)\hat{\mathcal{D}}dV.\label{vol}
\ee 
One can see that the form of volume transformation not only depends on the Doppler fact $\hat{\mathcal{D}}$, but also depends the Jacobian $J(v)$. Observationally, we are more interested in the specific flux $F_\nu$. For special relativity, the specific flux transformation is $F_{\nu'}'=\hat{\mathcal{D}}^3F_{\nu}$ for an isotropic point source\footnote{In this work, we are only interested in point sources. For extended sources, the specific flux transformation is $F_{\nu'}'=\hat{\mathcal{D}}F_{\nu}$ \citep{zha18}.} \citep{zha18}. Such a result is based on the results of volume transformation ($dV'=\hat{\mathcal{D}}dV$) and intensity transformation ($j_{\nu'}'=\hat{\mathcal{D}}^2j_{\nu}$), which results from Lorentz invariance \citep{der09}. Since Lorentz invariance is allowed to break in the RMS framework, similar to the volume transformation, the specific flux transformation can be written as
\be
F_{\nu'}'=G(v)\hat{\mathcal{D}}^3F_{\nu},\label{flux}
\ee
where $G(v)$ depends on the RMS parameters $a(v),b(v),d(v),\epsilon(v)$, and is a function of the relative velocity $v$.
$G(v)=1$ is the necessary condition for the Lorentz invariance. For the massive-photon theory, since the de Broglie-Proca theory is still under the framework of special relativity, Lorentz invariance remains valid so that one always has $G(v)=1$.

\section{Testing Lorentz invariance via spectral features}\label{lisf}

According to Section \ref{inv}, one can test special relativity by measuring a spectral feature of an object. First, we measure the Doppler factor by 
$\hat{\mathcal{D}}=\lambda/\lambda'$, where $\lambda$ and $\lambda'$ are the spectral-line frequency at the probe frame $K'$ and the Earth frame $K$, respectively. 
Next, we measure the ratio of the line fluxes in the two frames, i.e., $F_{\lambda'}'/F_{\lambda}$, where $F_{\lambda}$ and $F_{\lambda'}'$ can be taken as the specific fluxes at $\lambda$ and $\lambda'$, respectively. In order to measure the specific flux precisely, we make use of the continuum specific flux at the line center through continuum spectral fitting. According to Eq.(\ref{flux}), one finally gets
\be
G(v)=\left(\frac{F_{\nu'}'}{F_{\nu}}\right)\left(\frac{\nu}{\nu'}\right)^3=\left(\frac{F_{\lambda'}'}{F_{\lambda}}\right)\left(\frac{\lambda'}{\lambda}\right)^5,\label{gfun}
\ee
where $F_\nu d\nu=F_\lambda d\lambda$ is used in the second equation.
\begin{figure}[]
\centering
\includegraphics[angle=0,scale=0.45]{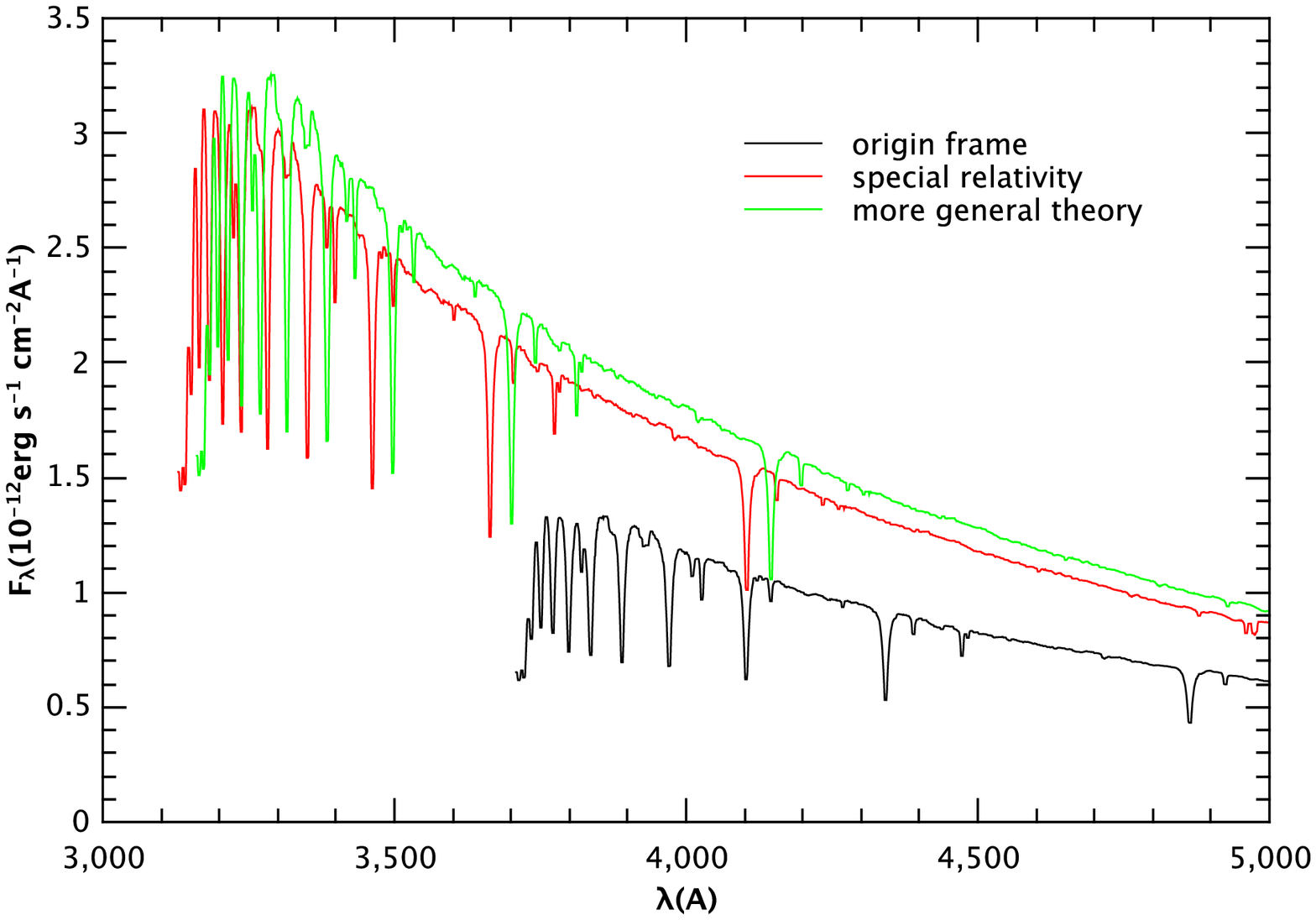}
\includegraphics[angle=0,scale=0.45]{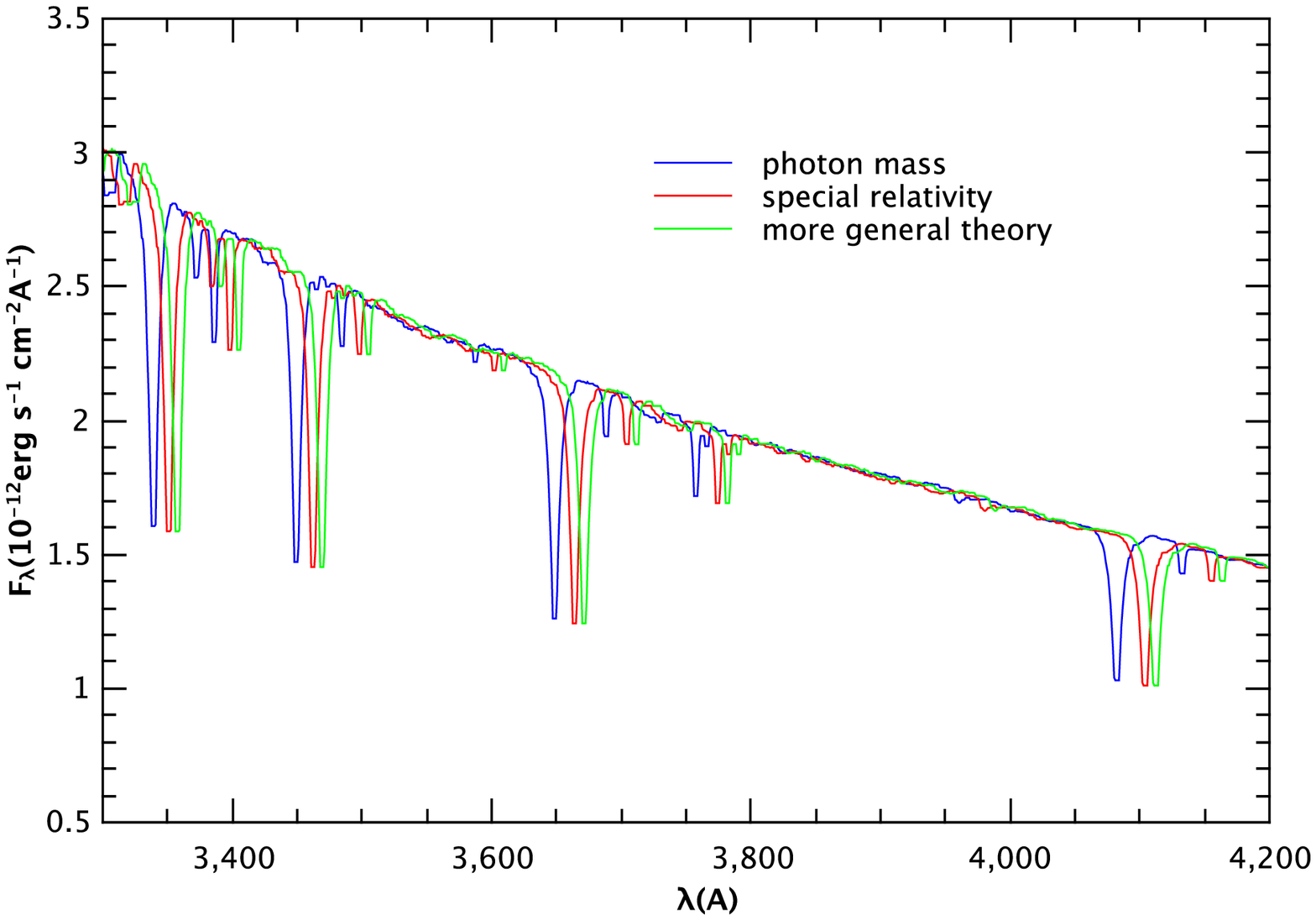}
\caption{Predicted spectra of HD 33688 (J051335.58+354201.7), an A-type star, as observed in the rest frame of the probe $K'$. The probe velocity is taken as $v=0.2c$ and the directional angle of the star from the direction of motion is taken as $\theta'=\pi/6$. Top panel: The black curve denotes the original spectrum in Earth frame $K$ (LAMOST archival data). The red curve denotes the spectrum in the probe frame for the case of special relativity. The green curve denotes the spectrum in the probe frame for the case of beyond special relativity with $\hat\gamma(v)=1+(\gamma-1)(1+50\%)=1.03093$ ($\gamma=1.02062$ for $\beta=0.2$) and $G(v)=1.1$. Bottom panel: The red curve denotes the spectrum in the probe frame for the special-relativity case. The blue curve denotes the spectrum in the probe frame with $m_\gamma=10^{-33}~\unit{g}$. The green curve denotes the spectrum in the probe frame for the beyond-special-relativity case with $\hat\gamma(v)=1+(\gamma-1)(1+10\%)=1.02268$ ($\gamma=1.02062$ for $\beta=0.2$) and $G(v)=1.01$.}\label{fig2}
\end{figure} 

In Figure \ref{fig2}, we simulate the spectra of an A-type star, HD 33688 (J051335.58+354201.7), in the probe frame $K'$ and Earth frame $K$. 
The observed spectrum of HD 33688 is from Large Sky Area Multi-Object Fiber Spectroscopic Telescope (LAMOST) archival data\footnote{http://dr.lamost.org} \citep{luo15}, which is denoted as the black curve. Assume that the probe velocity is $v=0.2c$, and that the angle between HD 33688 and the probe motion direction is $\theta'=\pi/6$. In the top panel of Figure \ref{fig2}, we simulate the optical spectral features of HD 33688. For the case of special relativity, the Doppler factor is $\mathcal{D}=1.18505$. The simulated spectrum is shown as the red curve. One can see that due to the Doppler effect, the spectrum is blue-shifted and the observed specific flux is amplified. If we use Eq.(\ref{gfun}) to test Lorentz invariance, one gets $G(v)\simeq1$ for the red curve, which agrees well with special relativity. 
Next, for the case beyond special relativity, we assume that\footnote{Here, $\hat\gamma(v)=1+(\gamma-1)(1+50\%)=1.03093$ means that $\hat\gamma-1$ has a $50\%$ deviation with respect to $\gamma-1$.} $\hat\gamma(v)=1+(\gamma-1)(1+50\%)=1.03093$ and $G(v)=1.1$ and simulate the corresponding spectrum, as shown by the green curve. One can see the green curve does not overlap with the red curve, although the probe parameters, e.g., $v$ and $\theta'$, are the same as before. Measuring $G(v)$ using Eq.(\ref{gfun}), one gets $G(v)\simeq1.1$, which agrees with the initial input value that slightly deviates from the prediction of Lorentz invariance, i.e., $G(v)=1$. 

In the bottom panel of Figure \ref{fig2}, we compare the spectra in the probe frame in the case of special relativity (without photon mass, red), the case of the massive-photon theory (blue) and a more generalized theory (beyond special relativity case, green). The red curve denotes the spectrum for the case of special relativity. The blue curve is the case with $m_\gamma = 10^{-33}$ g. The green curve denotes the case of more generalized theory beyond special relativity with $\hat\gamma(v)=1+(\gamma-1)(1+10\%)=1.02268$ ($\gamma=1.02062$ for $\beta=0.2$) and $G(v)=1.01$. 
Comparing both panels of Figure \ref{fig2}, in order to constrain $\Delta G(v)$ to the level of $\lesssim10\%$, the specific flux uncertainty $\Delta F_\lambda$ needs be less than $\sim10\%F_\lambda$, which is of the order of magnitude as the spectral fluctuation. Thus, the constraints on the Lorentz invariance, e.g., $G(v)$, is weak. On the other hand, in order to constrain $\Delta(\hat\gamma-1)\lesssim10\%$ by comparing with the  case of special relativity (i.e. the green vs. red curves in the bottom panel of Figure \ref{fig2}), the spectral resolution needs to satisfy $R\simeq\gamma/(\hat\gamma-\gamma)\gtrsim500$.  

Finally, for photon mass $m_\gamma=10^{-33}~\unit{g}$, as shown in the blue curve in the bottom panel of Figure \ref{fig2}, the corresponding specific flux is very close to that in the special-relativity case. However, since the Doppler factor $\mathcal{D}_m(\lambda')$ depends on the wavelength $\lambda'$, the shifts of the spectral lines at different wavelength is different from the classic Doppler shifts, see the lines around $335~\unit{nm}$ and $410~\unit{nm}$. Such a spectral feature can be used to constrain photon mass. We give an example.
In order to constrain photon mass $m_\gamma$, we choose three spectral lines with $\lambda_1'$, $\lambda_2'$ and $\lambda_3'$, respectively, and measure them in two different frames. According to Eq.(\ref{dopm}), one can solve $m_\gamma$ using the following formula:
\be
\frac{f_1/c_{\gamma2}'(m_\gamma) -f_2/c_{\gamma1}'(m_\gamma)}{f_1-f_2}=\frac{f_2/c_{\gamma3}'(m_\gamma) -f_3/c_{\gamma2}'(m_\gamma)}{f_2-f_3},\nonumber\\\label{sl}
\ee
where $c_{\gamma i}'(m_\gamma)\equiv c_\gamma'(\lambda_i',m_\gamma)$ and $f_i=\lambda_i'/\lambda_i$ with $i=1,2,3$, and the light speed $c_\gamma'(\lambda',m_\gamma)$ is given by Eq.(\ref{vm}). Using Eq.(\ref{sl}), one can solve for $m_\gamma$.
For example, we take the input values as $v=0.2c$, $\lambda_1'=350~\unit{nm}$, $\lambda_2'=400~\unit{nm}$, $\lambda_3'=450~\unit{nm}$, $\theta'=\pi/6$. 
For the spectral resolution of $R\sim(100-1000)$, according to Eq.(\ref{sl}), one can obtain the uncertainty of the photon mass as $\sigma_{m_\gamma}\sim10^{-33}~\unit{g}$. Since $c\geqslant c_\gamma'(\lambda')\simeq2.9979\times10^{10}~\unit{cm~s^{-1}}$ in the massive photon theory, the upper limit of photon mass would be constrained to the level $m_\gamma\lesssim10^{-33}~\unit{g}$. Notice that the above result is of the same order of magnitude as the mass defined by the optical frequency in the probe frame $m_\gamma<h/c\lambda'\sim6\times10^{-33}~\unit{g}$. Thus, such a constraint on the photon mass is very weak.

\section{Testing time dilation and photon mass via spectral line and directional angle}\label{slda}

In this section, we develop a method to test time dilation and photon mass via Doppler effect using the imaging and spectrograph capabilities of the transrelativistic probe.
First, we focus on the observational image distortions from astronomical objects. Again we take the Earth frame as $K$ and the probe frame as $K'$, and consider $N$ sources in each frame. According to the \emph{Theorem} in Appendix, the apparent displacement of a certain astronomical source in both frames is always towards  the direction or the anti-direction of the relative motion. Therefore, through comparing with the positions of $N$ sources in $K$ and $K'$ frame, see the algorithm of Appendix, one can obtain the motion direction. 
Notice that this method can be applied to theories beyond special relativity, since it only depends on symmetry.

We assume that the uncertainty of the sources in the probe frame is $\sim\lambda'/D$, where $D$ is the probe aperture. Using the method in the Appendix, we derive the uncertainty of the motion direction to be much less than $\sim\lambda'/D$ for the source number $N\gg1$ \cite[see also][]{zhu18}. In the following estimation, we take the uncertainty of the motion direction as of the order of $\sim\lambda'/D$ to be conservative\footnote{Notice that even if the uncertainty is larger, e.g. reaching $\sim1000\lambda/D$ (but $\ll\theta$), the following results essentially remain the same, because the uncertainty of the final results, e.g., $\Delta\gamma$ and $m_\gamma$, is dominated by the spectral resolution $R$ rather than the position uncertainty.}.

After determining the direction of motion, we chose two sources with their spectra measured, and define their directional angles related to the probe motion direction as $\theta_1'$ and $\theta_2'$, respectively, in the probe frame $K'$. According to Eq.(\ref{doppler}), one has
\be
\hat\gamma=\frac{\lambda_1'}{\lambda_1(1-\beta\cos\theta_1')}=\frac{\lambda_2'}{\lambda_2(1-\beta\cos\theta_2')},\label{twosource}
\ee
where $\lambda_1'$ and $\lambda_2'$ correspond to the wavelengths of the two spectral lines in the probe frame $K'$, and $\lambda_1$ and $\lambda_2$ correspond to the wavelengths of the same two spectral lines in the Earth frame $K$. Using Eq.(\ref{twosource}), one can obtain the probe dimensionless velocity as
\be
\beta=\frac{f_1-f_2}{f_1\cos\theta_2'-f_2\cos\theta_1'},\label{beta}
\ee
where $f_i=\lambda_i'/\lambda_i$ with $i=1,2$.
It is interesting to note that that such a derivation of the dimensionless velocity applies to more generalized theory beyond special relativity as well (see the detailed description in Section \ref{DE}). Substituting Eq.(\ref{beta}) into Eq.({\ref{twosource}}), one can obtain $\hat\gamma$. Define
\be
\Delta\hat\gamma\equiv\left|\hat\gamma-\gamma\right|=\left|\hat\gamma-\frac{1}{\sqrt{1-\beta^2}}\right|,
\ee
where $\gamma=1/\sqrt{1-\beta^2}$ is the Lorentz factor in special relativity. One can then test special relativity by directly measuring $\Delta\hat\gamma$ from the data. 

Once the probe motion velocity is obtained via this method, one can also constrain the photon mass under the framework of special relativity. For a given source with spectral lines, we measure its directional angle $\theta'$ in the probe frame $K'$, and also measure the wavelength $\lambda'$ and $\lambda$ of a certain spectral line in both $K'$ and $K$ frames, respectively. According to Eq.(\ref{dopm}), one can obtain the light speed $c_\gamma'(\lambda')$ in the probe frame, i.e.
\be
\frac{c_\gamma'(\lambda')}{c}=\beta\left(\frac{\gamma\lambda\cos\theta'}{\gamma\lambda-\lambda'}\right),
\ee
where $\beta$ is measured via Eq.(\ref{beta}) and $\gamma=1/\sqrt{1-\beta^2}$. Using Eq.(\ref{vm}), finally the photon mass is given by
\be
m_\gamma=\frac{h}{c\lambda'}\sqrt{1-\beta^2\left(\frac{\gamma\lambda\cos\theta'}{\gamma\lambda-\lambda'}\right)^2}.\label{mass}
\ee
In the massive photon theory, the ultimate speed must be larger than the light speed, i.e. $c\geqslant c_\gamma'(\lambda')\simeq2.9979\times10^{10}~\unit{cm~s^{-1}}$ in  a wide wavelength range, where the equal sign corresponds to $m_\gamma=0$. Thus, one can place an upper limit to the photon mass via Eq.(\ref{mass}).

At last, we analyze the uncertainty of the above method. Assume that the uncertainty of the directional angle of an object is about $\delta\theta\sim\lambda'/D$, where $D\sim3.5~\unit{cm}$ is the probe aperture for the Breakthrough Starshot. 
We take the input values as $v=0.2c$, $\lambda_1'=400~\unit{nm}$, $\lambda_2'=450~\unit{nm}$, $\theta_1'=\pi/6$ and $\theta_2'=\pi/4$. Considering the above uncertainties and using the Monte Carlo method, we simulate a sample of two sources with spectral lines. According Eq.(\ref{beta}), one can obtain the output value and the corresponding uncertainty. 
For a one-time measurement with the spectrograph resolution of $R\sim100$, the uncertainty of the probe motion speed is $\sigma_v\sim0.01c$, and the constraint on $\Delta\gamma$ could reach $\Delta\gamma\lesssim0.01$. If the probe spectrograph can reach $R\sim1000$, one would have $\sigma_v\sim0.001c$ and $\Delta\gamma\lesssim0.001$.  
In the standard Robertson-Mansouri-Sexl (RMS) framework \citep{rob49,man77}, the parameter $\alpha$ satisfies $|\alpha+1/2|=|1/\hat\gamma-1/\gamma|/\beta^2\simeq\Delta\gamma/\gamma^2\beta^2$. Thus, one has $|\alpha+1/2|\lesssim0.25$ for $R\sim100$, and $|\alpha+1/2|\lesssim0.025$ for $R\sim1000$.
Finally, based on Eq.(\ref{mass}), we can also constrain the mass of photon. For the above parameters and uncertainty, the uncertainty of the photon mass upper limit is $\sigma_{m_\gamma}\sim10^{-33}~\unit{g}$, leading to $m_\gamma\lesssim10^{-33}~\unit{g}$, which weakly depends on $R$. 

\section{Discussion and Conclusion}\label{discon}

When a probe travels at a good fraction of speed of light, some interesting relativistic effects would occur \citep[e.g.][]{pen59,ter59,chr17}. Among many possible observations one could perform with a transrelativistic camera \citep[][]{zha18}, testing the Doppler effect would be one interesting and important task to carry out.

In this work, we propose to use the Doppler effect to test special relativity and constrain photon mass. Through introducing a generalized time dilation factor $\hat\gamma(v)$, we obtain the form of the generalized Doppler factor $\hat{\mathcal{D}}$ which has the same form as the classic form, see Eq.(\ref{doppler}). For theories beyond special relativity, Lorentz invariance, including invariant four-volume $dtd^3\bm{x}=dtdV$, the invariant phase-space element $d^3\bm{p}/E$ and the invariant phase volume $d\mathcal{V}=d^3\bm{x}d^3\bm{p}$, may break. In this case, more generalized Doppler transformations, 
e.g. Eq.(\ref{vol}) and Eq.(\ref{flux}), should be introduced. 

On the other hand, the massive-photon theory is still within the framework of special relativity. According to the classical energy-momentum relation, the lower the photon frequency, the slower the light velocity. Replacing the ultimate maximum speed $c$ with the photon speed $c_\gamma'(\nu')$, one can obtain the frequency Doppler effect relation, i.e., $\nu'=\mathcal{D}_m(\nu')\nu$, with the Doppler factor $\mathcal{D}_m(\nu')$ being the function of the photon frequency. Since the massive-photon theory is within the framework of special relativity, Lorentz invariance, including invariant four-volume, invariant phase-space element and invariant phase volume, remain valid. As a result, other Doppler transformations e.g, volume ($dV'=\mathcal{D}_mdV$), intensity ($j_{\nu'}'=\mathcal{D}_m^2j_{\nu}$) etc., also keep the same forms as the classic ones, with $\mathcal{D}_m$ replaced by $\mathcal{D}$.

Based on the above theories, one can perform a test of special relativity and constrain the photon mass by observing the spectral features and images of one or more sources within the rest frame of a transrelativistic probe. Comparing the results with those obtained from Earth, one can carry out many interesting constraints. First, based on the information of the probe motion derived from the imaging method \citep{zhu18} and the measured spectral information, one can compare theoretically derived $\mathcal{D}_{\rm the}$ and observationally measured $\mathcal{D}_{\rm obs}$ to directly test the prediction from special relativity (Section \ref{zhu}). Next, for generalized theories beyond special relativity, the specific flux Doppler relation becomes $F_{\nu'}'=G(v)\hat{\mathcal{D}}^3F_{\nu}$. One can use the spectral information only to measure $G(v)$ to test the Lorentz invariance (Section \ref{lisf}). We show that in order to constrain $\Delta G(v)\lesssim10\%$, the specific flux uncertainty $\Delta F_\lambda$ needs be less than $\sim10\%F_\lambda$, which is of the order of magnitude as the spectral fluctuation. On the other hand, in order to constrain $\Delta(\hat\gamma-1)\lesssim10\%$ by comparing with the special-relativity case, the spectral resolution needs to be $R\gtrsim500$. Next, we show that the photon mass can be constrained by three spectral lines of a certain astronomical object measured in both the probe and the Earth frames (Section \ref{lisf}). The upper limit of the photon mass can be constrained to $m_\gamma\lesssim10^{-33}~\unit{g}$. Finally, by combining the imaging information and spectral information, one can apply a different method to perform tests for the above theories (Section \ref{slda}). 
For a probe traveling with velocity $v\sim0.2c$, aperture $D\sim3.5~\unit{cm}$ and spectral resolution $R\sim100$, the probe velocity uncertainty can be constrained to $\sigma_v\sim0.01c$ and the time dilation factor uncertainty can be constrained to $\Delta\gamma=|\hat\gamma(v)-\gamma|\lesssim0.01$ which corresponds to $|\alpha+1/2|\lesssim0.25$ in the RMS framework. 
If the probe spectral resolution is allowed to be higher, e.g., $R\sim1000$, one would have $\sigma_v\sim0.001c$ and $\Delta\gamma\lesssim0.001$, which corresponds to $|\alpha+1/2|\lesssim0.025$ in the RMS framework.
On the other hand, once the probe velocity is obtained, we can further constrain the photon mass using a certain spectral line, deriving a mass upper limit of $m_\gamma\lesssim10^{-33}~\unit{g}$.

Testing special relativity can be performed by laboratory experiment methods \citep[e.g.][]{rei07,her09,tor09} and astrophysical methods \citep[e.g.][]{zha18,zhu18} on different aspects of special relativity. In this work, we propose that some astrophysical methods allow direct measurements of the Doppler factor from spectral features and imaging to perform a self-consistency check. Since Doppler effect is directly associated time dilation in special relativity (see Section \ref{DE}), 
through sending a probe moving at an appreciable fraction of the speed of light, time dilation can be constrained by measuring spectral features and imaging of astronomical objects in different frames. Once Breakthrough Starshot is launched in the future, the spectral features of astronomical objects could be used as the new astrophysical probe to test time dilation, although the current design parameter cannot provide a better constraint on time dilation than that in the recent Ives–-Stilwell experiments with $|\alpha+1/2|\lesssim10^{-8}$ \citep{rei07,bot14}.

Similar to the aberration method in \citet{zhu18}, the above methods can also constrain the photon mass not too much below the photon energy in the optical band. Thus, these methods of constraining the photon mass are not competitive compared with other astrophysical methods, e.g., the frequency dependence of the speed of light \citep{lov64,wu16,sha17,wei18,xin19}, the spindown of pulsars \citep{yan17}, the solar wind magnetic field \citep{ryu97,ryu07,ret16}, etc.

\acknowledgments 
We thank the anonymous referee for helpful comments and suggestions.
The authors acknowledge Su Yao for providing the Large Sky Area Multi-Object Fiber Spectroscopic Telescope (LAMOST) stellar spectra. 
The LAMOST is operated and managed by the National Astronomical Observatories, Chinese Academy of Sciences. 

\appendix
\section{Algorithm to solve the motion direction}

\begin{figure}[H]
\centering
\includegraphics[angle=0,scale=0.35]{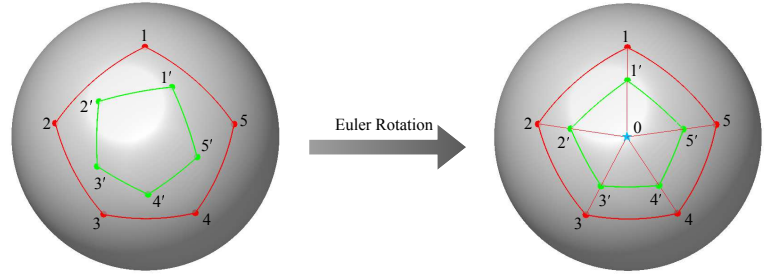}
\caption{Solving the probe motion direction via Euler rotation.}\label{fig3}
\end{figure}

For the RMS theory that is beyond special relativity, one can obtain the probe motion direction directly according to the following theorem\footnote{In \citet{zhu18}, the motion velocity and direction of the probe can be solved within the framework of special relativity. For the RMS theory, according to this theorem, one can obtain the probe motion direction directly, but the motion velocity is unknown in this case.}:
\begin{itemize}
\item \emph{Theorem}: Due to \emph{symmetry}, the position of a source in the observer frame $K$ is \emph{always} in the plane defined by its position in the comoving frame $K'$ and the direction of the relative motion. Therefore, in the celestial coordinate system, the apparent displacement of a certain astronomical source between the two frames is always towards the direction or the anti-direction of the relative motion.
\end{itemize}

Next, we describe an algorithm to solve the motion direction according to the \emph{Theorem}. First, we define three basic Euler rotation matrices to rotate vectors by an angle $\theta$ around the $x$-, $y$-, or $z$-axis using the right-hand rule:
\begin{eqnarray}
R_x(\theta)=
\begin{pmatrix}
1 & 0 & 0\\
0 & \cos\theta & -\sin\theta\\
0 & \sin\theta & \cos\theta\\
\end{pmatrix},
~~~
R_y(\theta)=
\begin{pmatrix}
\cos\theta & 0 & \sin\theta\\
0 & 1 & 0\\
-\sin\theta & 0 & \cos\theta\\
\end{pmatrix},
~~~
R_z(\theta)=
\begin{pmatrix}
\cos\theta & -\sin\theta & 0\\
\sin\theta & \cos\theta & 0\\
0 & 0 & 1\\
\end{pmatrix}.
\end{eqnarray}

\begin{enumerate}

\item Label the objects in the Earth frame as $1,2,3...N$ and the corresponding objects in the probe frame as $1',2',3'...N'$, as shown in Figure \ref{fig3}, and define their coordinates as $(x_i,y_i,z_i)$ and $(x_{i'}',y_{i'}',z_{i'}')$ for $i=1,2,3...N$, respectively. Since all points are in the sphere, one always has $x^2+y^2+z^2=1$. 

\item Take any two pairs of the corresponding points, e.g., $(x_i,y_i,z_i)$, $(x_{i'}',y_{i'}',z_{i'}')$ and $(x_j,y_j,z_j)$, $(x_{j'}',y_{j'}',z_{j'}')$. Their great-circle formulae must satisfy the forms
\be
x^2+y^2+(A_ix+B_iy)^2-1&=&0,\label{gcirc1}\\
x^2+y^2+(A_jx+B_jy)^2-1&=&0,\label{gcirc2}
\ee
respectively, where $A_i$ and $B_i$ depend on $(x_i,y_i,z_i)$ and $(x_{i'}',y_{i'}',z_{i'}')$, and $A_j$ and $B_j$ depend on $(x_j,y_j,z_j)$ and $(x_{j'}',y_{j'}',z_{j'}')$. Using Eq.(\ref{gcirc1}) and Eq.(\ref{gcirc2}), one can solve two intersections. We only leave the one with $z>0$ and define it as $(x_{ij},y_{ij},z_{ij})$, which depends on $(A_i,B_i)$ and $(A_j,B_j)$.

\item Based on the above information, one can obtain $N(N-1)/2$ intersections $(x_{ij},y_{ij},z_{ij})$ with $i,j=1,2,3...N$ and $i\neq j$. Take any two intersections, e.g., $(x_{ij},y_{ij},z_{ij})$ and $(x_{mn},y_{mn},z_{mn})$, to calculate their spherical distance $d_{ijmn}$, which is defined as the ``intersection distance''. There are $(N^4-2N^3-N^2+2N)/8$ intersection distances. At last, one can set the maximum intersection distance as 
\be
s=\max(\{d_{ijmn}\}),
\ee
where $\{d_{ijmn}\}$ is the list of the intersection distances.

\item Rotate the probe frame via
\be
R_z(\theta_1)R_x(\theta_2)R_z(\theta_3)
\left(
\begin{array}{c}
x\\
y\\
z\\
\end{array}
\right)
\rightarrow
\left(
\begin{array}{c}
x\\
y\\
z\\
\end{array}
\right),
\ee
and repeat steps 1 to 4, as shown in Figure \ref{fig3}. Take $0\leqslant\theta_1,\theta_2,\theta_3\leqslant 2\pi$ with an angle step $\delta \theta_i\rightarrow0$ with $i=1,2,3$, we obtain a list of $\{s_i\}$. The motion direction would correspond to the limit of the intersections when $s_i\rightarrow0$. 

\end{enumerate}

\end{document}